\documentclass[a4paper,11pt]{article}
\pdfoutput=1 

\usepackage{jcappub} 

\usepackage{aas_macros}
\usepackage{natbib}
\usepackage{amsfonts}
\usepackage{amsmath}
\usepackage{amssymb}
\usepackage{graphicx}
\usepackage[dvipsnames]{xcolor}
\usepackage{hyperref}
\hypersetup{colorlinks=true,allcolors=teal}
\usepackage{caption}
\usepackage{subcaption}

\newcommand{\Msun}{\ensuremath{M_{\odot}}}
\newcommand{\Mh}{\ensuremath{h^{-1}M_{\odot}}}

\newcommand{\Mpch}{\ensuremath{h^{-1}{\rm Mpc}}}

\newcommand{\be}{\begin{equation}}
\newcommand{\ee}{\end{equation}}
\newcommand{\Cal}[1]{\ensuremath{\mathcal{#1}}}

\usepackage{multirow}
\usepackage{tabularx}

\newcommand{\Om}{\ensuremath{\Omega_{\rm m}}}
\newcommand{\Ob}{\ensuremath{\Omega_{\rm b}}}
\newcommand{\Ok}{\ensuremath{\Omega_{\rm k}}}

\newcommand{\ns}{\ensuremath{n_{\rm s}}}
\newcommand{\As}{\ensuremath{A_{\rm s}}}

\newcommand{\Vpeak}{\ensuremath{V_{\rm{peak}}}}
\newcommand{\Sinhagad}{\texttt{Sinhagad}}
\newcommand{\Sahyadri}{\texttt{Sahyadri}}
\usepackage[normalem]{ulem}

\title{Stabilizing simulation-based cosmological Fisher forecasts: a case study using the Voronoi volume function}

\author[a,1]{Saee Dhawalikar,\note{Corresponding author.}}
\author[a]{Aseem Paranjape,}
\author[b]{and Shadab Alam}
\affiliation{Inter-University Centre for Astronomy \& Astrophysics,\\ Ganeshkhind, Post Bag 4, Pune 411007, India}
\affiliation{Tata Institute of Fundamental Research, Homi Bhabha Road, Mumbai 400005, India }

\emailAdd{saee.dhawalikar@iucaa.in}
\emailAdd{aseem@iucaa.in}
\emailAdd{shadab.alam@tifr.res.in}

\abstract{Forecasting cosmological constraints from halo-based statistics often suffers from instability in derivative estimates, especially when the number of simulations is limited. This instability reduces the reliability of Fisher forecasts and machine learning based approaches that use derivatives. We introduce a general framework that addresses this challenge by stabilizing the input statistic and then systematically identifying the optimal subset of summary statistics that maximizes cosmological information while simultaneously minimizing the instability of predicted constraints. 
We demonstrate this framework using the halo mass function as well as the Voronoi volume function (VVF), a summary statistic that captures beyond two-point clustering information. Applying our two-step procedure -- random sub-sampling followed by optimization -- improves the constraining power by up to a factor of 4, while also enhancing the stability of the forecasts across realizations.
As surveys like Euclid, DESI, and LSST push toward tighter constraints, the ability to produce stable and accurate theoretical predictions is essential. Our results suggest that new summary statistics such as the VVF, combined with careful data curation and stabilization strategies, can play a key role in next-generation precision cosmology.}


\begin{document}
\maketitle
\flushbottom

\section{Introduction}
\label{sec:Introduction}
The Large-Scale Structure (LSS) of the Universe contains a wealth of information about the underlying cosmological model (including the nature of dark matter, dark energy and initial conditions) as well as the physics of galaxy formation and evolution. With the increasing volume, depth, and precision of ongoing and upcoming LSS surveys such as the Dark Energy Spectroscopic Instrument (DESI) \citep{DESI_2016_a, DESI_DR2_Results_II_2025}, Rubin-LSST \citep{LSST2019} and Euclid \citep{Euclid2011}, it is now possible to probe the matter distribution of the Universe with unprecedented detail. These surveys will unlock immense information; at the same time they present a significant analysis challenge: identifying summary statistics that are both robust and maximally informative for constraining cosmological parameters.

Traditionally, analyses of the LSS have relied on two-point statistics like the two-point correlation function (2PCF) and the power spectrum, which would represent complete summaries if cosmic structure were Gaussian distributed. However, as structure formation becomes increasingly nonlinear at late times, a significant fraction of the cosmological information resides in higher-order correlations that are not captured by these conventional methods. This motivates the exploration of alternative and complementary summary statistics that can probe non-Gaussian features and environmental dependence more effectively.

In recent years, a wide range of non-Gaussian, halo-based statistics have been developed to capture such higher-order information. These include, but are not limited to, the k-nearest-neighbor (kNN) distribution \citep{KNN2021_a, KNN2021_b}, marked correlation functions \citep{Marked_correlations_sheth+2004, White2016} and Minkowski functionals \citep{Minkowski_functionals1997, Minkowski_SDSS2003}. Each of these probes the LSS in a unique way, targeting higher-order spatial correlations. In addition, simulation-based inference (SBI) methods that utilize machine learning have emerged, which employ neural networks to learn compact summary statistics directly from simulations \citep{Alsing&Wandelt2018, Charnock+2018, SBI_BNN2023, SBI_CNN2024, Semelin+2025}.  The Voronoi volume function (VVF), introduced in \citep{VVF2020} and used below, is another recent promising summary statistic, that accesses the beyond-Gaussian information through the distribution of cell volumes in the Voronoi tessellation of the given set of tracers. As demonstrated in recent studies \citep{VVF2020,VVF2024}, VVF captures higher-order clustering and assembly bias signatures, and potentially provides cosmological constraints that are complementary to those from the 2PCF.

Despite the potential of these advanced halo-based statistics, their use in Fisher forecasts and cosmological inference faces practical challenges. The estimation of robust covariance matrices and stable numerical derivatives of these statistics with respect to cosmological parameters -- which is typically done via finite differences using simulation suites -- requires a large number of simulation realizations. 
For instance, robust Fisher analyses and simulation-based inference applications often allocate $\Cal{O}(100)$ realizations specifically for derivative estimation \citep{Hahn+2020_Fisher, Banerjee+2022_KNN, Alsing+2019, Makinen+2022}.
This is not always feasible, especially in case of high-resolution large-volume simulations. Without careful treatment, this limitation leads to noisy derivatives and unreliable Fisher estimates. The need for a robust, reproducible and computationally efficient method to extract information from such halo-based statistics is therefore imminent. 

In this paper, we present a general formalism that reduces the noise in the choice of summary statistic by averaging over multiple random sub-samples, and then identifies a subset of data points that maximizes the stability and information content in the Fisher forecast. Our method is agnostic to the specific choice of statistic, and relies on a general method to stabilize the statistic of interest using random sub-sampling and simple diagnostics to quantify the informativeness, accuracy and stability of derivatives with respect to the cosmological parameters to be constrained, without invoking expensive machine learning techniques. While the halo mass function (HMF) and VVF serve as a case study in this paper, our method is general and applicable to any halo-based, field-based or neural summary statistic involving noisy derivative estimates. As we enter an era of precision cosmology driven by increasingly complex datasets, such methods will be essential to extract reliable constraints from the nonlinear Universe.

The paper is organized as follows. Section~\ref{sec:Numerical_techniques} describes the suite of N-body simulations used in this work, along with the tracer sample selection. It also provides a brief overview of the VVF and outlines how it is used for Fisher forecasting. In Section~\ref{sec:Statistics}, we present our formalism for selecting an optimal subset of data points for Fisher analysis in two scenarios: (a) when multiple realizations are available to estimate the covariance matrix, along with a few variations of a single parameter at a time providing an accurate derivative estimate ; and (b) when only a single high resolution realization per cosmology is available with only a 3-point finite difference derivative estimate, and splitting in sub-boxes is required for estimating the covariance. In Section~\ref{sec:Results}, we apply this method to the VVF and the HMF measured from different tracer samples and demonstrate its effectiveness, and conclude in Section~\ref{sec:Discussion}.

\section{Numerical Techniques}
\label{sec:Numerical_techniques}
\begin{table}[h!]
    \centering
    \begin{tabular}{|c|c|c|}
    \hline
        Parameter & Default value ($\theta_0$) & Variation magnitude $\Delta $  \\
        \hline
        \Om &  0.3138 & 0.05 $\Omega_{m0}$ \\
        \ns &  0.9649 & 0.05 $n_{s0}$ \\
        $h$ &  0.6736 & 0.05 $h_0$ \\
        \As &  2.0989$\times 10^{-9}$ & 0.1 $A_{s0}$ \\
        \Ob &  0.0493 & 0.1 $\Omega_{b0}$ \\
        \Ok &  0.0 & 0.05 \\
        \hline
    \end{tabular}
    \caption{Cosmological parameters and variations of the simulation suites.}
    \label{tab:cos_params}
\end{table}

\begin{table}[h!]
\centering
    \begin{tabular}{|c|c|c|c|c|}
    \hline
    Parameter & \multicolumn{2}{c|}{Variation magnitude} & \multicolumn{2}{c|}{Realizations} \\
    \cline{2-5} 
    & \Sinhagad & \Sahyadri &\Sinhagad & \Sahyadri  \\
    \hline
    Default & - & - & 101 &1\\
    \Om  &  $\pm 2 \Delta, \pm \Delta, \pm \Delta/2$ & $\pm \Delta$ & 10 &1 \\
    \ns  &  $\pm 2 \Delta, \pm \Delta, \pm \Delta/2$ & $\pm \Delta$ & 10 &1 \\
    $h$  & $\pm 2 \Delta, \pm \Delta$ & $\pm \Delta$ & 1 &1 \\
    \As  & $\pm 2 \Delta, \pm \Delta$ & $\pm \Delta$ & 1 &1 \\
    \Ob  & $\pm 2 \Delta, \pm \Delta$ & $\pm \Delta$ & 1 &1 \\
    \Ok & $\pm 2 \Delta, \pm \Delta$ & $\pm \Delta$ & 1 &1 \\
    \hline
    \end{tabular}
    \caption{Summary of number of simulations in the \Sinhagad\, and \Sahyadri\, simulation suites for each cosmological parameter variation.}
    \label{tab:sinhagad_summary}
\end{table}

\subsection{N-body simulations}
\label{subsec:Simulations}
We use two suites of N-body simulations named \Sinhagad\, and \Sahyadri\, to perform Fisher forecast analysis based on the HMF and the VVF. \Sinhagad\, serves as a pilot study for the higher-resolution \Sahyadri\ simulation suite. The default simulations adopt cosmological parameters consistent with the Planck 2018 results \citep{Planck18}. For the Fisher analysis, each of the six cosmological parameters is varied individually while keeping the others fixed. Table~\ref{tab:cos_params} summarizes the default parameter values ($\theta_0$) along with their respective variations ($\Delta$).

All simulations have a periodic comoving box of size $L_{\rm{box}}= 200\,\Mpch$ and $256^3$ ($2048^3$) particles, corresponding to a particle mass of $4.15 \times 10^{10}\, (8.1 \times 10^7)\Mh$ in case of the \Sinhagad\, (\Sahyadri) suite. In the \Sinhagad\, suite, for each cosmological parameter, $4$ simulations are performed, with the parameters taking values $\theta_0 \pm 2\Delta, \theta_0 \pm \Delta$. These variation simulations are seed matched with one of the default realizations, namely $r1$. Additionally, nine more sets of seed matched simulation variations are performed for $\Om, \ns$, along with variations of $\pm \Delta/2$. These two parameters  are the primary focus for this suite in this paper. For the estimation of the covariance matrices, 100 additional realizations of the default simulation are used. 
In the \Sahyadri\ suite (described in detail by \citep{Sahyadri}), two simulations are performed for each cosmological parameter at $\theta_0\pm \Delta$, also seed matched to $r_1$. For this suite, we shall focus on the parameters $\Om, h$. A summary of all simulations in both the suites is provided in Table~\ref{tab:sinhagad_summary}.

Simulations in both the suites are performed using the tree-PM code \textsc{gadget-4} \citep{Gadget4}, with a comoving force softening length set to $1/30$ of the mean inter-particle spacing, and a PM grid of $512^3$ (\Sinhagad) and $4096^3$ (\Sahyadri). Initial conditions are generated at redshift $z = 49$ using second-order Lagrangian Perturbation Theory \citep{Scoccimarro1998}, implemented via \textsc{N-GenIC}, which is integrated into \textsc{gadget-4}. A total of 201 (\Sinhagad) and 101 (\Sahyadri) snapshots are saved between $z = 12$ and $z = 0$, evenly spaced in scale factor $a \equiv (1 + z)^{-1}$, and are used for constructing halo catalogs and merger trees (see below). All the analysis in this work, however, is performed using the snapshot at $z=0$ in each simulation.

To obtain multiple realizations for both the derivative estimates and the covariance calculations, we subdivide each simulation volume into smaller sub-boxes and treat these as independent realizations. Each \Sinhagad\ box is partitioned into $2^3$ cubical sub-boxes, and each \Sahyadri\ box into $3^3$ sub-boxes. This yields $27$ realizations for both the seed-matched derivative estimates and the covariance matrix in the \Sahyadri\, suite. For \Sinhagad, it results in $80$ realizations for each derivative estimate and $800$ realizations for the covariance calculation.

All simulations and analyses used in this paper were carried out on the Pegasus cluster at IUCAA, Pune.\footnote{\url{http://hpc.iucaa.in}}

\subsection{Halo samples}
\label{subsec:halo_samples}
Dark matter halos in the simulations are identified using the six-dimensional phase-space Friends-of-Friends algorithm implemented in \textsc{rockstar} \citep{Rockstar1023}, and halo merger trees are constructed with the \textsc{consistent-trees} code \citep{Consistent_trees2013}. Halo samples are selected based on a threshold in their maximum circular velocity along the main progenitor branch in the merger tree, denoted as \Vpeak. This quantity is chosen because it closely correlates with the stellar mass of the associated galaxy in subhalo abundance matching (see, e.g., \citep{SHAM_2015, SHAM2016, SHAM_RSD2022}). Only halos resolved with at least 40 particles are considered, with halo mass defined as $m_{200\rm{b}}$. Here, $m_{200\rm{b}}$ refers to the gravitationally bound mass within a radius $R_{200\rm{b}}$, which encloses a region with density 200 times the mean matter density of the Universe. Additionally, halos are cleaned based on the degree of relaxation -- $\eta \equiv 2T/|U|$, where $T$ and $U$ denote the total kinetic and gravitational potential energy of the halo, respectively. Following \citep{Bett+2007}, only halos with $0.5 \leq \eta \leq 1.5$ are retained. The selection thresholds are chosen to maintain a fixed number density of tracers across all simulations for each sample. 

Choosing a very high number density introduces sample incompleteness, as a larger fraction of halos fall below the resolution threshold of 40 particles when selecting by \Vpeak. To minimize this, we choose the number density such that, before enforcing the 40 particle cut, the fraction of unresolved halos in the sample selected by the corresponding \Vpeak\, cut is $\leq 1\%$ of the total sample.
Conversely, very low number densities lead to increased noise in both the statistic measurement (see section~\ref{subsec: Numerical_VVF} for details) and the corresponding covariance matrix estimates. Balancing these considerations, two samples are selected with comoving number densities of $n = 2 \times 10^{-2}$ and $2 \times 10^{-3}\,(\rm{Mpc})^{-3}$ for the \Sahyadri\, suite and $n=2 \times 10^{-4} (\rm{Mpc})^{-3}$ for the \Sinhagad\, suite.

\subsection{Voronoi volume function}
\label{subsec: Numerical_VVF}
\begin{figure}
\centering
\includegraphics[width=\linewidth]{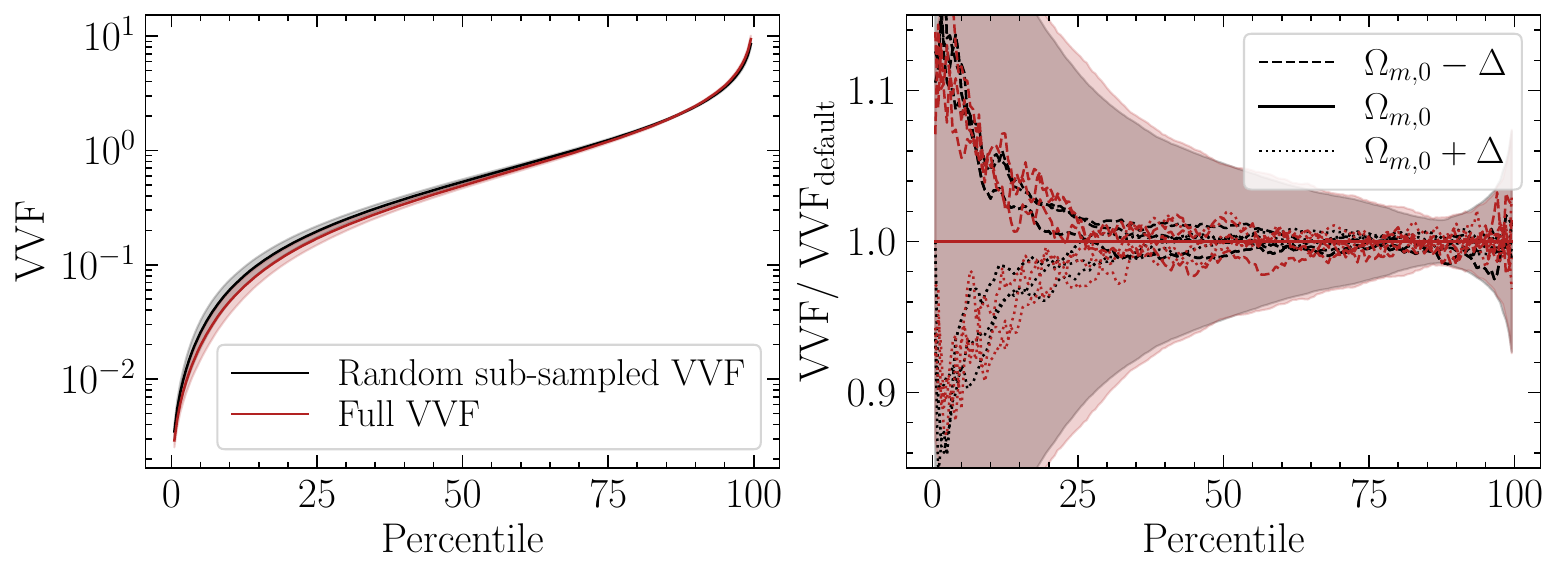}
\caption{Illustration of VVF of the highest number density ($n=2\times 10^{-2}\,\rm{Mpc}^{-3}$) tracer sample from the \Sahyadri\, simulation suite for different sub-boxes and choice of summary statistics. \emph{(Left panel):} VVF calculated using the full tracer sample in red, and the one obtained by averaging over random sub-samples in black. The solid line shows the mean VVF, and the bands show the diagonal errors from the covariance matrices. \emph{(Right panel):} Random sub-sampled (black) and full VVF (red) for three random sub-boxes, with variations in $\Om$. Each curve is divided by the VVF for the default simulation in the corresponding sub-box. Dashed (dotted) curves show the simulations with lower (higher) $\Om$. Shaded regions show the diagonal errors, same as the left panel. It is seen that even after seed matching, the VVF variations are unstable. See text for details.}
\label{fig:VVF_intro}
\end{figure}

The Voronoi volume function (VVF), introduced in \citep{VVF2020}, is defined as the cumulative distribution of cell volumes in the Voronoi tessellation \citep{Voronoi1908} constructed from a set of points, such as the locations of galaxies or halos, within a given volume. Each cell represents the region closest to a particular point, with its volume inversely proportional to the local tracer number density. By capturing higher-order correlations in the spatial distribution of tracers, the VVF provides information beyond traditional two-point statistics like the two-point correlation function (2PCF). This makes it a potentially powerful probe of the non-Gaussian features of large-scale structure, which are sensitive to both cosmological parameters and galaxy formation physics.

As a relatively new statistic in cosmology, the VVF was employed alongside the 2PCF in \citep{VVF2024} to model the redshift-space distribution of galaxies in the GAMA survey \citep{GAMA2011, GAMA2022}. This combined analysis enabled a tentative detection of assembly bias -- where galaxy properties depend on environmental factors beyond the host halo mass. The VVF’s sensitivity to such higher-order environmental effects highlights its potential as a valuable tool for jointly constraining cosmology and galaxy formation models. While the primary focus of this paper is not to assess the constraining power of the VVF itself, we use it here to demonstrate the applicability of the general formalism we develop for robust and stable parameter inference.

The VVF is computed following the method described in \cite{VVF2020} with an addition: the use of random sub-sampling to suppress noise and improve stability.
For each tracer sample, 70\% of the tracers are randomly selected without repetition, and the VVF is computed. This procedure is repeated 10 times, and the final VVF is obtained by averaging over these 10 realizations. Unless stated otherwise, this averaged VVF is the default statistic throughout the paper (see Appendix~\ref{App: bootstrp vs full VVF} for the effect of the random sub-sampling and sensitivity of the results on the sub-sampling choice.) The VVF is evaluated over percentiles ranging from 0.5 to 100, in steps of 0.5. The \emph{left panel} in Figure~\ref{fig:VVF_intro} compare the VVF obtained using the full sample to that from the averaged sub-sample realizations for the highest number density sample in \Sahyadri.  While sub-sampling naturally alters the shape of the VVF due to the reduced sampling density (see Appendix~B1 of \citep{VVF2020} for some theoretical insights), it also leads to a smoother and more stable estimate, making it the preferred choice for the subsequent analysis. However, we have found that this improvement is prominent only for lower number density samples, where the VVF estimates are more noisy. As one moves to higher number densities, the effect of random sub-sampling becomes less pronounced. The \emph{right panel} shows how the VVF varies with changes in \Om. We discuss this further in the next section.

\section{Statistical techniques}
\label{sec:Statistics}
The goal of any summary statistic is to retain maximal information about cosmological parameters in as compact a form as possible. In principle, if the data were optimally compressed -- as, e.g., in the linear scheme MOPED \citep{MOPED2000} -- one would only need as many summary values as there are parameters. In practice, however, due to model non-linearities and/or parameter-dependent covariances, one might be forced to use a larger number of summary data points. At the same time, the number of summary data points used in the Fisher forecast -- denoted by $N_d$ --  must be limited. This is because estimating the covariance matrix of the statistic across realizations becomes increasingly noisy and unstable as the size of the data vector grows -- unless a very large number of simulations is available. The objective, therefore, is to select an optimal subset of $N_d$ data points from the full dataset such that the information content is maximized while ensuring that the predicted constraints remain stable across different realizations. 

To illustrate the need for such optimization, the \emph{right panel} of Figure~\ref{fig:VVF_intro} shows the VVF computed from different seed-matched simulations under small variations in the cosmological parameter $\Om$, using both the full tracer sample (red) and random sub-samples (black). Despite using seed-matched initial conditions, large variations persist across realizations, leading to noisy and unreliable numerical derivatives. These fluctuations directly impact the accuracy and consistency of Fisher forecasts, highlighting the need for a principled method to select a subset of stable and informative statistics.

Below, we introduce an optimization framework based on three statistical measures, minimizing which allows us to quantify the balance between information gain and stability.
We provide two versions of this framework: one suited for scenarios where a reliable estimate of the true parameter derivatives are available, and another applicable when such data are not accessible. The former is applied to the \Sinhagad\, suite, while the latter is suitable for use with the \Sahyadri\, simulations.

\subsection{\Sinhagad-like setup}
\label{subsec:stats_sinhagad}
Here, we want to focus on statistics that are sensitive to non-Gaussian information, and can work with slightly smaller volumes. To balance the advantages of working with larger effective volumes against the requirement of having enough realizations for covariance estimation and stable numerical derivatives, we subdivide each simulation into smaller sub-boxes and treat each sub-box as an independent realization. In case of \Sinhagad, for each of the cosmological parameters considered, we have $\pm 2\Delta, \pm \Delta, \pm \Delta/2$ variations available along with the default. Thus, accurate derivative estimate is obtained by first fitting a third order polynomial through the 7 points available at each VVF percentile from the seed-matched variation simulations, and using its first order coefficient as the estimate of the accurate derivative, $(dm/d\theta)_0$. The noisy slope estimate, which we denote $dm/d\theta$, is obtained via a three-point central difference using the $\pm \Delta$ variations. We use the accurate and noisy estimates together to quantify the stability of numerical derivative estimates.

To quantify the contribution of each data point to the Fisher information, we define:
\begin{equation}
\label{eq:Y}
Y_i \equiv \log{\left(\frac{\sigma_i}{\langle (dm_i/d\theta)_{0,k} \rangle_k} \right)},
\end{equation}
where $\sigma_i \equiv \sqrt{C_{ii}}$for the $i$-th data point or summary; the subscript $k$  is the index of the simulation realization and the average is taken over these realizations. Smaller values of $Y_i$ indicate that the $i$-th data point carries more information.

To evaluate the accuracy and stability of the derivative estimates, we define the ratio $R^k_i$ between the noisy and accurate derivative estimates for the $i$-th data point in the $k$-th realization:
\begin{equation}
\label{eq:R}
R_i^k \equiv \left[\frac{dm_i/d\theta}{(dm_i/d\theta)_0}\right]_k.
\end{equation}
Ideally, $R_i^k = 1$. We then define $X_i$, which measures the bias of the noisy slope estimate relative to the accurate slope:
\begin{equation}
\label{eq:X}
X_i \equiv \left| \log_{10}{\langle R_i^k \rangle_k} \right|, \\
\end{equation}
The averages of $R_i^k$ are calculated across the realizations. Finally, to quantify the variability of the slope across realizations, we define:
\begin{equation}
    \label{eq:Z}
    Z_i  \equiv \frac{\sigma_k((dm_i/d\theta)_{0,k})}{\left|\langle (dm_i/d\theta)_{0,k} \rangle_k\right|}
\end{equation}
where $\sigma_k$ denotes the standard deviation across the realizations.
A data point is retained only if $\langle R_i^k \rangle_k \geq 0$. In this regime, the logarithm defining $X_i$ is well defined, $X_i \geq 0$ and $Z_i \geq 1$, with $X_i = 0$ and $Z_i = 1$ corresponding to the ideal case.
To select data points that yield both stable derivatives and high information content, we impose a threshold $Z_{\rm{th}}$ and define a selection criterion in the $X$–$Y$ plane. For each cosmological parameter, the valid data points are those that (i) satisfy $Z_i \leq Z_{\rm th}$ for all parameters, and (ii) lie below the designated line in the $X$–$Y$ plane. The values of $Z_{\rm th}$ and the line parameters are chosen empirically, and their selection is discussed in Section~\ref{sec:Results}, where the procedure is applied using the VVF and HMF. Appendix~\ref{App: bootstrp vs full VVF} shows that using the more stable, randomly sub-sampled VVF yields a larger set of valid points, underscoring the benefit of random sub-sampling.

If the number of valid data points, $N_v$, exceeds the desired number $N_d$, a subset of size $N_d$ must be chosen for the Fisher analysis. Assuming neighboring points are more strongly correlated — true in the case of VVF -- it is advantageous to select points that are as uncorrelated as possible. While the optimal selection would ideally account for the full covariance structure, doing so is computationally expensive. As a simple proxy that encourages de-correlation, we generate all possible equally spaced sets of $N_d$ points from the $N_v$ valid points and perform Fisher forecasts for each.

To evaluate the stability of the resulting forecasts, we compare the Fisher-based Gaussian posterior distribution from each realization with the one obtained by averaging the inverse Fisher matrices of each individual realization using the Kullback-Leibler divergence ($D_{\rm{KL}}$) \citep{KL_div}, which for Gaussian distributions reduces to:
\begin{equation}
\label{eq:KL}
    D_{\rm{KL}} (p_1 \lVert p_2)=\frac{1}{2\ln{2}}\left[\mathrm{Tr}(\Sigma_2^{-1}\Sigma_1)-d-\ln\mathrm{Det}({\Sigma_2^{-1}\Sigma_1})+(\mu_2-\mu_1)^T\Sigma_2^{-1}(\mu_2-\mu_1)\right]
,\end{equation}
where $\Sigma \equiv F^{-1}, \mu$ are the inverse covariance matrix and mean of the parameter posterior distribution, and $d$ is dimension, which in our case is $2$. The first distribution is the expected distribution, with the corresponding covariance matrix $\Sigma_2$, which is $\langle F^{-1}_k \rangle_k$ in this case. In our case, $\mu_2=\mu_1$, which is equal to the parameter values in the fiducial model. Thus, the last term is always zero. The $D_{KL}$ value is quoted in bits. We calculate the average $D_{\rm{KL}}$ across all realizations for each subset of points. The optimal subset of $N_d$ points is the one that minimizes this average KL divergence. Although we adopt an equally spaced sampling for computational efficiency, this is not a strict requirement.  In cases where correlations between neighboring points are weak, an exhaustive search over all combinations of $N_d$ points from the $N_v$ valid ones may be performed.

\subsection{\Sahyadri-like setup}
\label{subsec:stats_Sahyadri}
In the absence of availability of an accurately estimated slope, we modify $Y$ by replacing the quantity $\langle (dm_i/d\theta)_0 \rangle$ with $\langle dm_i/d\theta \rangle_{k}$, the average of the three-point finite-difference derivative computed across the equivalent realizations:
\begin{equation}
\label{eq:Ys}
    Y_i^s \equiv \log{\left(\frac{\sigma_i}{\langle (dm_i/d\theta)_k \rangle_{\rm{k}}} \right)}\,.
\end{equation}
$X$ (equation~\ref{eq:X}) quantifies the accuracy of the derivative in a single realization, and $Z$ (equation~\ref{eq:Z}) quantifies its stability across realizations. Thus, we modify these quantities to be:
 \begin{align}
\label{eq:Rs}
R_{i,k}^s& \equiv \frac{\max(|S_{i, k}|)}{\min(|S_{i,k}|)}\, \\
\label{eq:Xs}
X_i^s & \equiv  \log_{10}{\langle R_{i,k}^s \rangle_k}, \\
\label{eq:Zs}
Z_i^s &\equiv \frac{\sigma_{k} (dm_i/d\theta)}{|\langle  (dm_i/d\theta)_k  \rangle_{k}|},
\end{align}
To ensure that we only retain points whose slope estimates are consistent in sign within a given realization, we define another quantity:
\begin{equation}
\label{eq:Qs}
    Q_i^s \equiv  \left \langle \frac{\max(S_{i,k})}{\min(S_{i,k})} \right \rangle_{k}.
\end{equation}
The set $S_i=\{(dm_i/d\theta)_b,\, (dm_i/d\theta), \, (dm_i/d\theta)_f \}$  includes the backward, central, and forward finite-difference slope estimates, computed using $\pm \Delta$ variations. The $\max, \min$ in equations~\ref{eq:Xs}, \ref{eq:Qs} are taken over this set of 3 values for each data point indexed by $i$, for the $k$-th realization.

After computing these metrics, we first discard those points with $Q_i^s \le 0$. This is the logical equivalent of discarding points with $\langle R_i^k \rangle_k\leq0$ in the earlier analysis. Then, we treat $X^s, Y^s, Z^s$ in equations~\eqref{eq:Ys}-\eqref{eq:Zs} to be conceptually equivalent to $X, Y, Z$ from equations~\eqref{eq:Y}-\eqref{eq:Z}, and follow the same procedure to identify the optimal subset of points based on a combination of a cut in the $X^s-Y^s$ plane, and a cut on $Z^s$; followed by minimization of KL divergence. Note that, again, the specific choices of thresholding are application-specific.

\section{Results}
\label{sec:Results}
This section demonstrates the application of the formalism developed above by using the HMF and VVF to constrain two cosmological parameters: $\Om$ ,$\ns$ for the \Sinhagad\ suite, and $\Om$, $h$ for the \Sahyadri\ suite. The choice of $\Om$ and $\ns$ for \Sinhagad\ is motivated by their impact on the amplitude and scale dependence of large-scale structure, to which the VVF is particularly sensitive. For \Sahyadri, the parameters $\Om$ and $h$ are selected as the full set of simulations required for these variations are available only for these at present. We apply the method to different tracer samples and assess the robustness and effectiveness of the proposed optimization procedure.

In Section~\ref{subsec: Results HMF}, we apply the optimization procedure to the HMF using the \Sahyadri\ simulations; the \Sinhagad\ resolution is insufficient to yield reliable HMF constraints. Section~\ref{subsec: Results VVF} extends the analysis to the VVF, where we apply the same optimization scheme and examine the stability of the resulting Fisher forecasts using tracer samples drawn from both simulation suites. 
\subsection{Halo Mass Function}
\label{subsec: Results HMF}
Here, we focus on a widely used statistic, the halo mass function defined to be:
\begin{equation}
     \rm{MF} \equiv \frac{d N}{d \log_{10} M}
\end{equation}
where $N, M$ are the number density of halos and the center of each mass bin in units of $\rm{Mpc}^{-3}, \Msun$ respectively. We choose logarithmically spaced mass bins with $d\log_{10}M=0.1$, and the mass, HMF are measured in units of $\Msun, \,\rm{Mpc}^{-3}$ respectively. For large enough volumes, we expect the resulting bin counts to be approximately correlated \citep{Hu&Kravstov2003}. Hence, here we use a diagonal covariance matrix.

We follow the procedure outlined in Section~\ref{subsec:stats_Sahyadri} to calculate the $X^s$, $Y^s$, and $Z^s$ statistics for both cosmological parameters of interest and for the two tracer samples. For the mass function, we find that random sub-sampling does not offer any advantage. Therefore, here we present the results with the full sample, without random sub-sampling.

\begin{figure}
\centering
\begin{subfigure}[b]{0.49\textwidth}
    \centering
    \includegraphics[width=\textwidth]{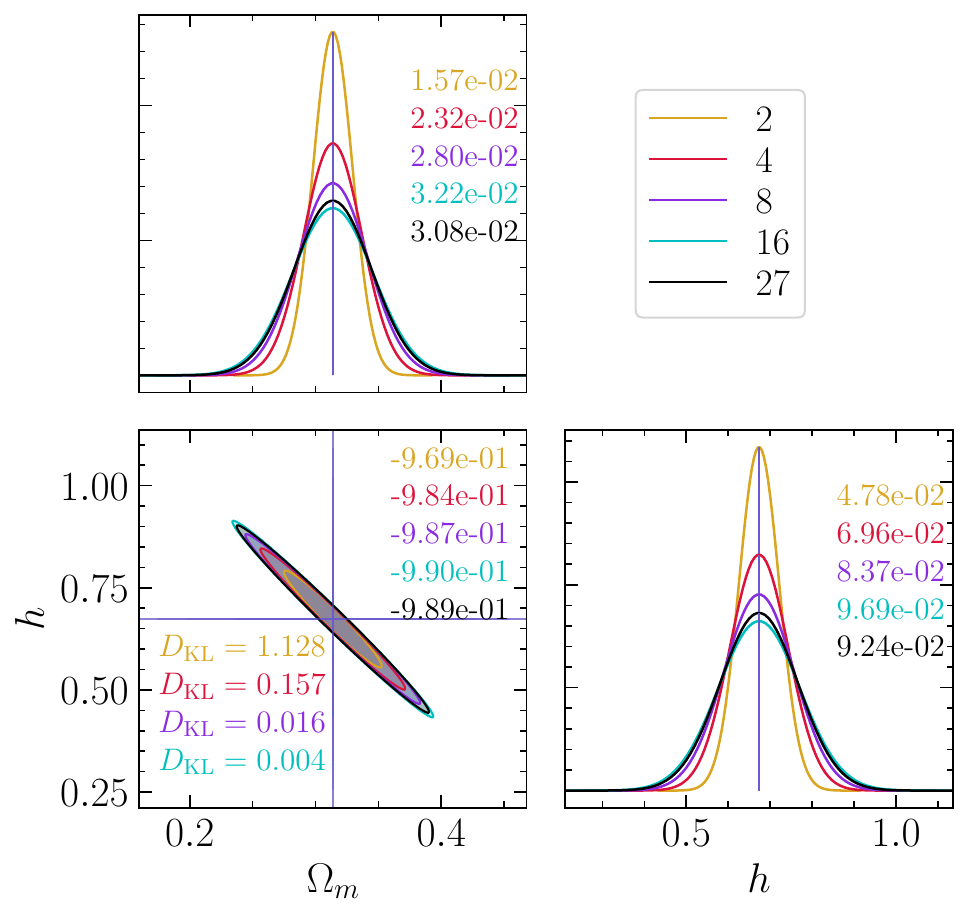}
    \caption{$n=2\times 10^{-2}$, unoptimized}
    \label{fig:corner_org}
\end{subfigure}
\hfill
\begin{subfigure}[b]{0.49\textwidth}
    \centering
    \includegraphics[width=\textwidth]{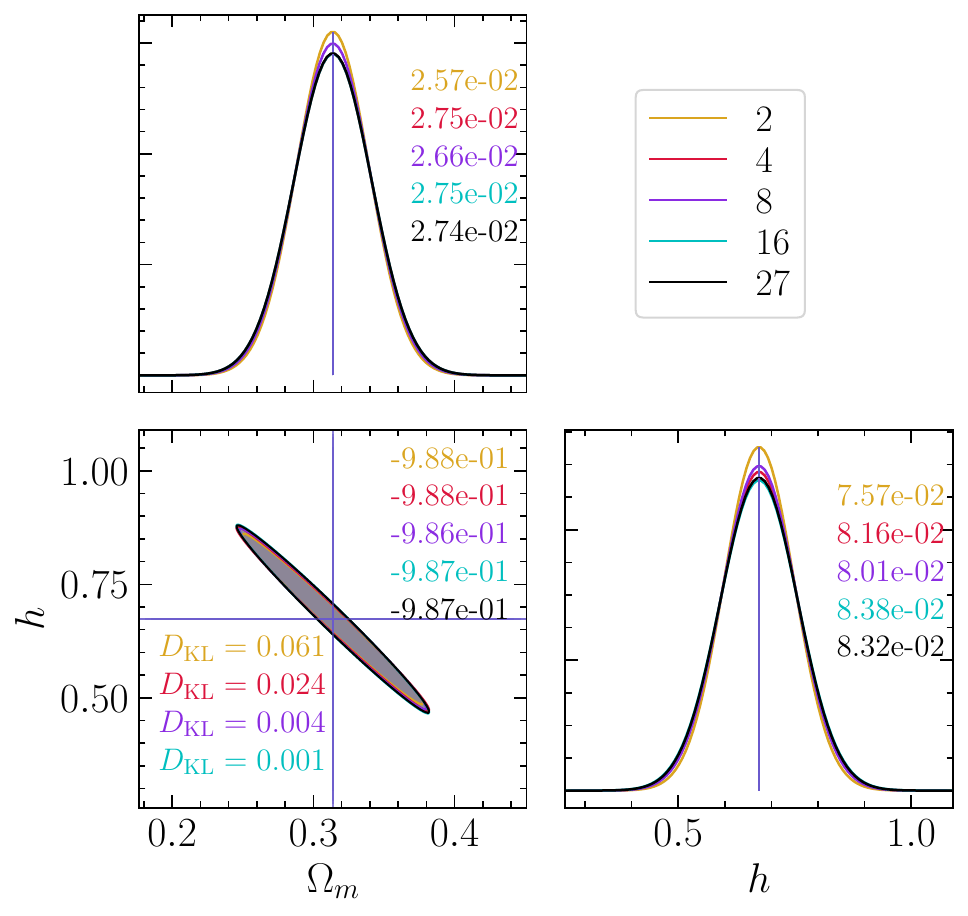}
    \caption{$n=2\times 10^{-2}$, optimized}
    \label{fig:corner_bs_org}
\end{subfigure}
\hfill
\begin{subfigure}[b]{0.49\textwidth}
    \centering
    \includegraphics[width=\textwidth]{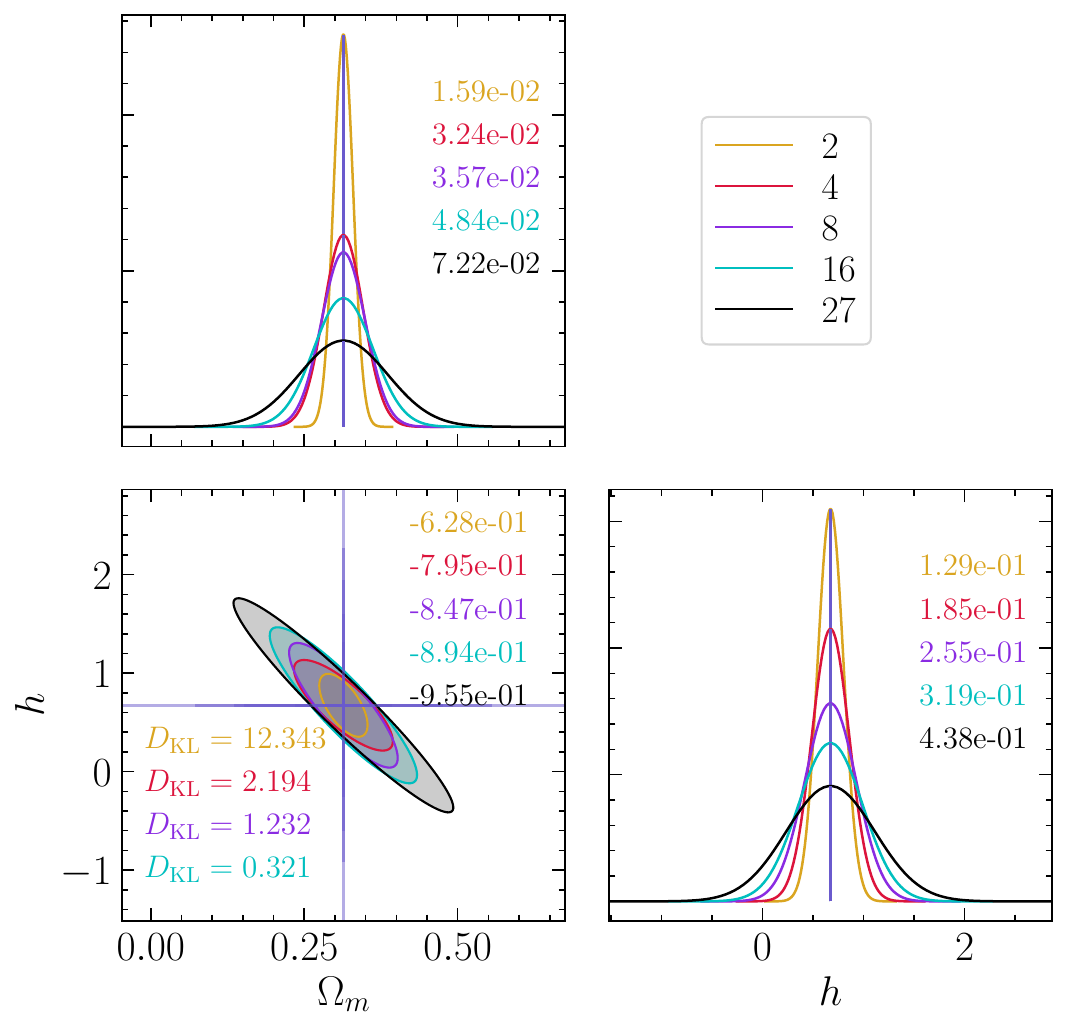}
    \caption{$n=2\times 10^{-3}$, unoptimized}
    \label{fig:corner_full_opt}   
\end{subfigure}
\begin{subfigure}[b]{0.49\textwidth}
    \centering
    \includegraphics[width=\textwidth]{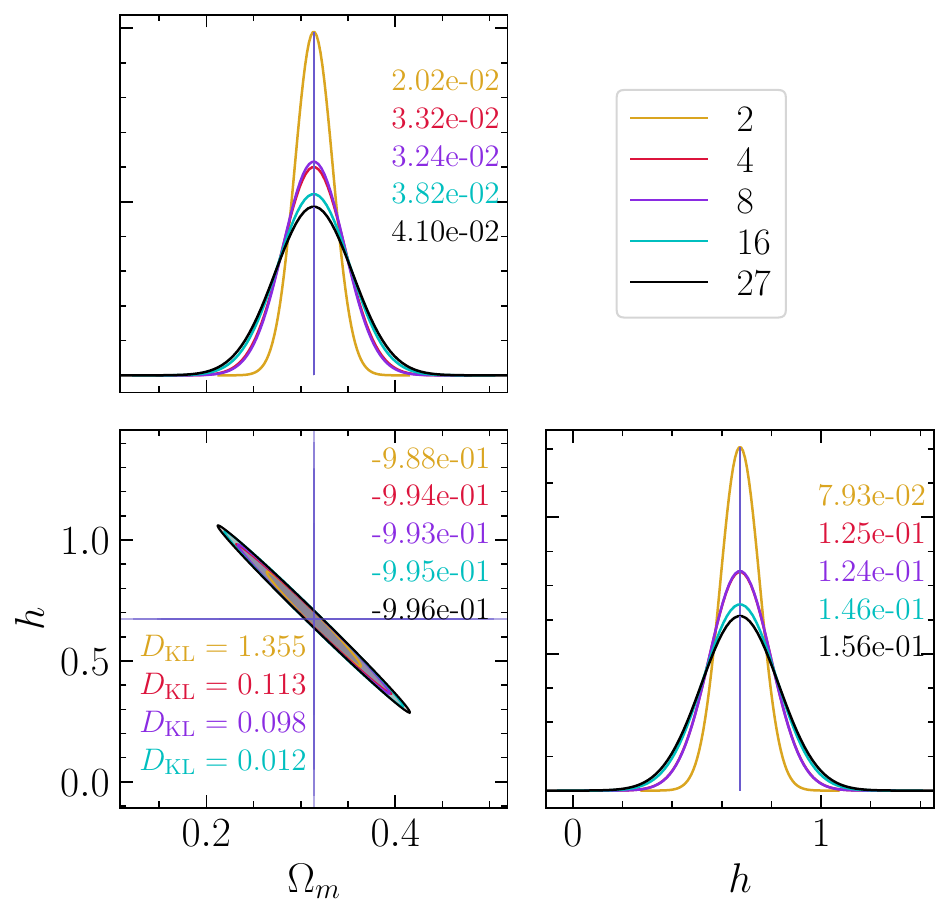}
    \caption{$n=2\times 10^{-3}$, optimized}
    \label{fig:corner_final}  
\end{subfigure}
\caption{Corner plots from Fisher analysis performed on the two tracer samples from \Sahyadri\, using HMF with and without optimization. The ellipses show $95.4 \%$ confidence regions. \emph{(Top (Bottom) panels)} show results obtained using the higher (lower) number density sample. \emph{Left (right) panels} show results without (with) optimization. Black colour shows the true Fisher constraints obtained by averaging the derivatives across all $27$ realizations. Yellow, red, purple and cyan show one random realization of the constraints obtained using averaging across $2, 4, 6,8$ realizations respectively. The widths ($\sigma$) of the marginalized 1D distribution are quoted in corresponding colours along with the 1D distributions. The correlation coefficients are quoted alongside the 2D ellipses in the upper right corners. The KL divergence (in bits) calculated with respect to the corresponding truth is reported in the bottom corner. It is seen that optimization improves both the constraining power and stability of the results. }
\label{fig:mf_fisher}
\end{figure}

\begin{figure}
    \centering
    \includegraphics[width=\linewidth]{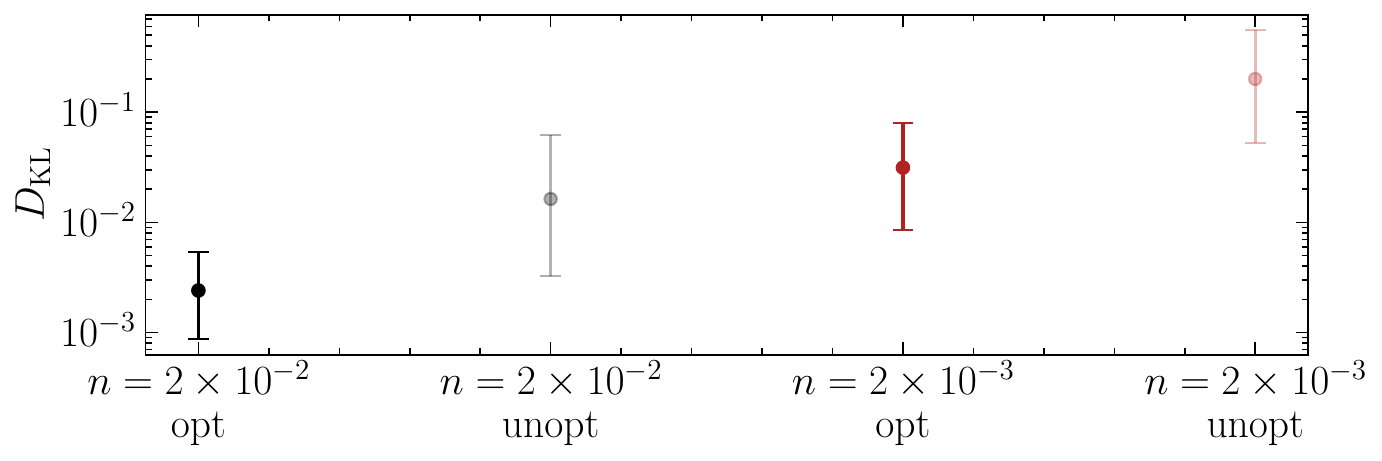}
    \caption{Comparison of Fisher constraints from multiple combinations of $16$ realizations with the accurate prediction for the mass function. Markers show the median value of KL divergence (in bits) with error bars showing the $16$th and $84$th percentiles. Darker markers show the optimized results and the lighter versions show the unoptimized counterparts. The optimized results show consistently lower KL divergence.}
    \label{fig:mf_kl}
\end{figure}

For the higher and lower number density samples considered, we impose thresholds of $Z^s_{\rm{th}}=1, 2$ respectively. We find that this threshold along with the $Q^s \geq 0$ condition is sufficient to optimize the data; the selected bins naturally correspond to low values of $X^s, Y^s$, and no additional cuts in that plane are necessary. The results are not strongly sensitive to the precise choices of $Z^s_{\rm{th}}$. These cuts lead to selection of $13, 7$ mass bins respectively, and no further sub-selection is required.

Because a single realization yields biased constraints, we average the derivatives over progressively larger numbers of realizations before computing the Fisher matrix. The right-hand panels of Figure~\ref{fig:mf_fisher} show the behavior for one random ordering of realizations: the constraints converge within $4–8$ realizations for both number densities, with the KL divergence reaching $\le 0.1$ bits. For comparison, we repeat the analysis using randomly selected mass bins (matching the optimized sample sizes). The corresponding results, shown in the left-hand panels, converge much more slowly as the number of realizations increases, and for the lower-density sample show no clear convergence even at 16 realizations. Also, note that the size of the constraints has shrunk substantially, especially for the lower number density sample, with the improvement in the marginalized errors on $h$ improving by a factor $2.8$.

To quantify convergence, we draw $10000$ random combinations of $16$ realizations each and compute the KL divergence between the resulting Fisher constraints and the accurate constraint. Figure~\ref{fig:mf_kl} shows the median and associated error bars for both tracer samples, comparing the optimized selection with the unoptimized case. The optimized samples consistently yield lower values of $D_{\rm{KL}}$, demonstrating the improved stability and robustness of the method.

\subsection{Voronoi Volume Function}
\label{subsec: Results VVF}

\begin{figure}
    \centering
    \includegraphics[width=\linewidth]{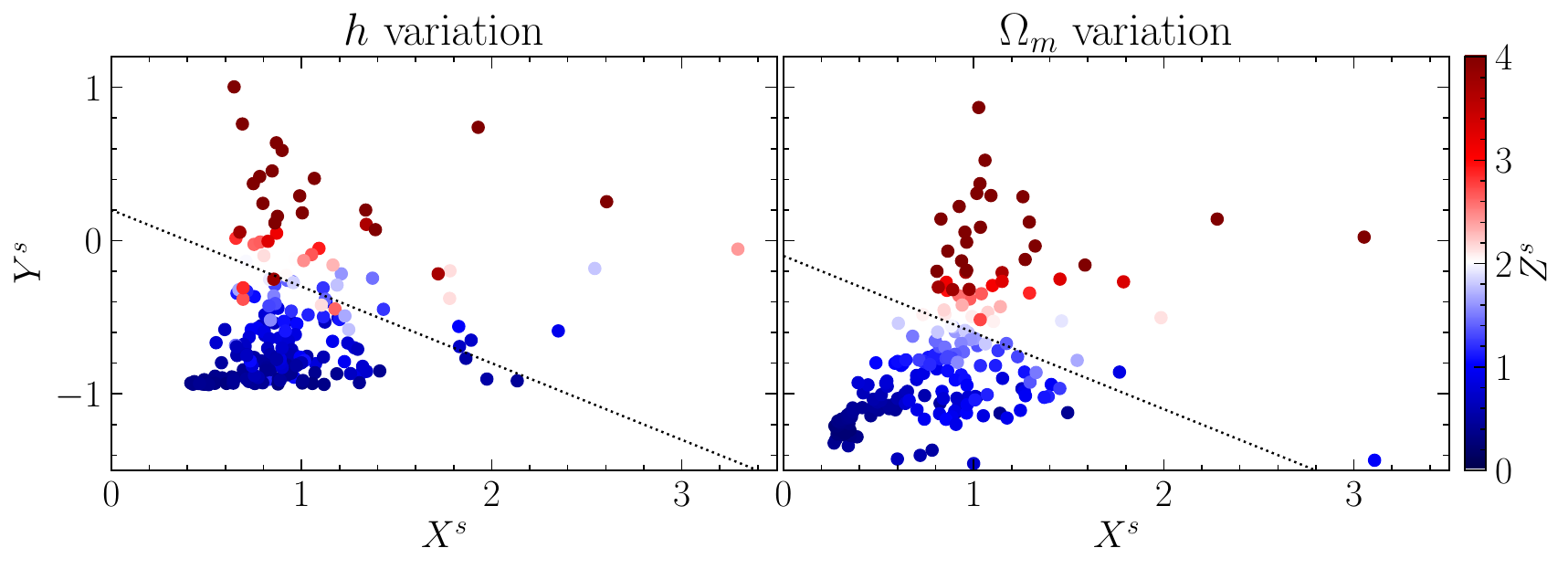}
    \caption{Scatter plots of the $X^s$ and $Y^s$ statistics for the highest number density tracer sample and two cosmological parameters. Each point represents a VVF data point, colour-coded by its $Z^s$ value, with $Z_{\rm{th}}=2$. Only points lying below the dotted lines are considered as valid data points. See Section~\ref{subsec: Results VVF} for further details.}
    \label{fig:2d_bootstrap}
\end{figure}

Here, we follow the procedure explained in sections~\ref{subsec:stats_sinhagad}, ~\ref{subsec:stats_Sahyadri} for one number density in \Sinhagad\ and two in \Sahyadri, respectively. 
For the highest to lowest number density, we choose $Z_{\rm{th}}=2, 5, 1.3$ respectively. Figure~\ref{fig:2d_bootstrap} shows the distribution of the points in the $X^s-Y^s$ plane. One can observe that there is a strong clustering of low-$Z^s$ points in the lower left corners, and a single line in each panel can approximately separate these low-$Z$ points from the rest. This is chosen visually to separate the strongly clustered solid points in the left bottom corner of each panel from the other scattered, high-$Z^s$ points. We select only those data points that (a) lie below this line for both cosmological parameters, and (b) satisfy $Z^s \leq Z_{\rm{th}}$, for a given number density sample.

\begin{figure}
     \centering
     \includegraphics[width=\linewidth]{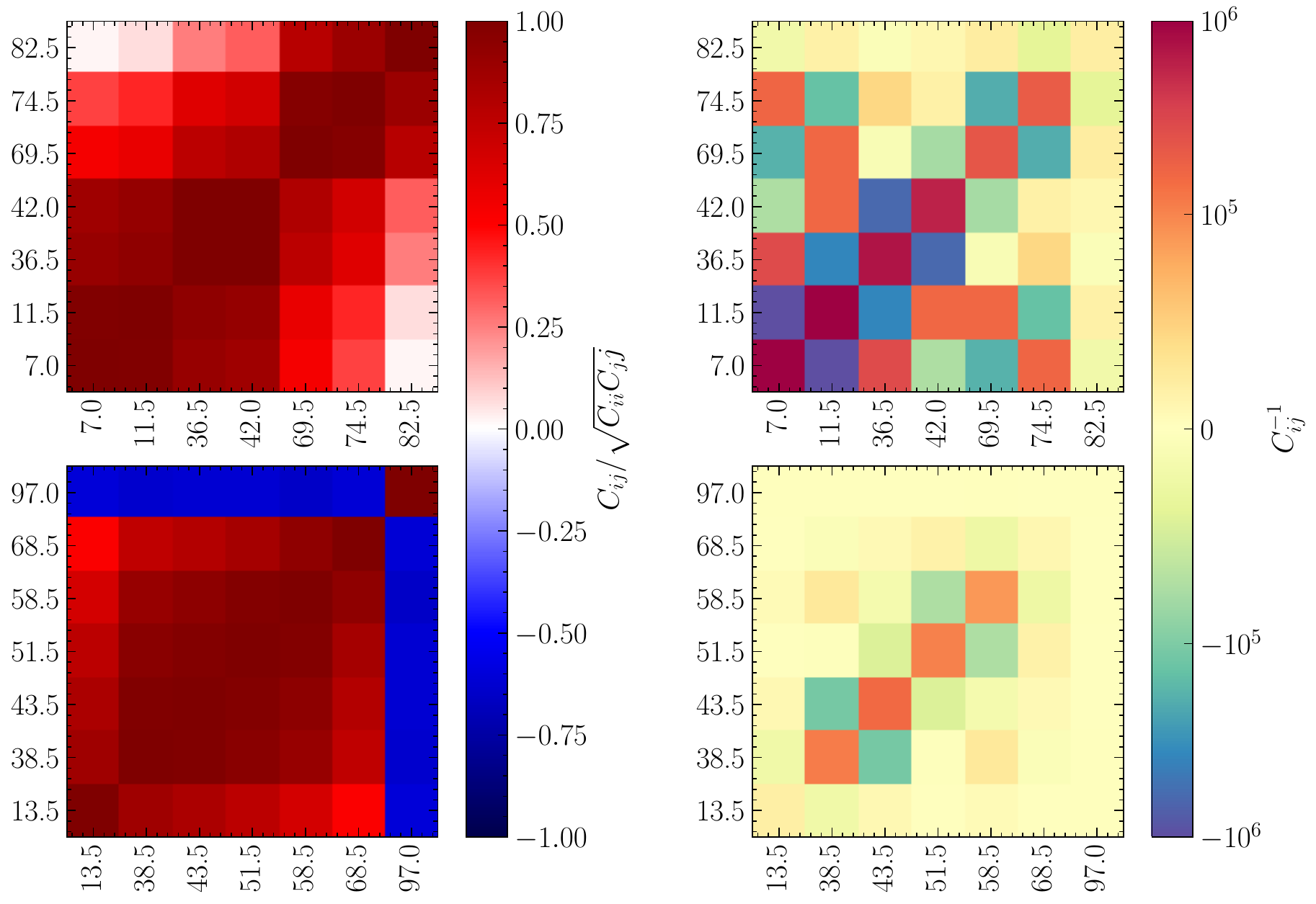}
     \caption{Normalized covariance (left) and inverse covariance matrices (right) for the two number density samples $n=2\times 10^{-2}$ (top) and $2\times 10^{-3} $ (bottom), for the optimized percentiles. It is seen that there is strong correlation between the percentiles, the higher percentiles being negatively correlated with the lower ones. These correlations are mostly limited to only neighboring percentile bins for the inverse correlation matrices, especially for the lower density sample.}
     \label{fig:covariance}
 \end{figure}

The behavior of the other two samples is qualitatively similar to the highest density one. This leads to a total of $55, 28,  26$ points in the three number density samples respectively. For the \Sahyadri\ samples, we apply our final optimization procedure using the KL divergence to obtain $7$ percentiles for each tracer sample. This number is constrained by the available $27$ realizations to estimate the covariance ($(7\times 8) /2=28$).
 The normalized and inverse covariance matrices corresponding to these percentiles are shown in Figure~\ref{fig:covariance}. We note that lower percentiles (specifically, those $\leq 13.5$ and $\leq 7.0$ for the higher and lower density tracer populations, respectively) are typically excluded. Additionally, there is a strong anti-correlation between the highest and lower percentiles.

\begin{figure}[h!]
     \centering
     \begin{subfigure}[b]{0.49\textwidth}
         \centering
         \includegraphics[width=\textwidth]{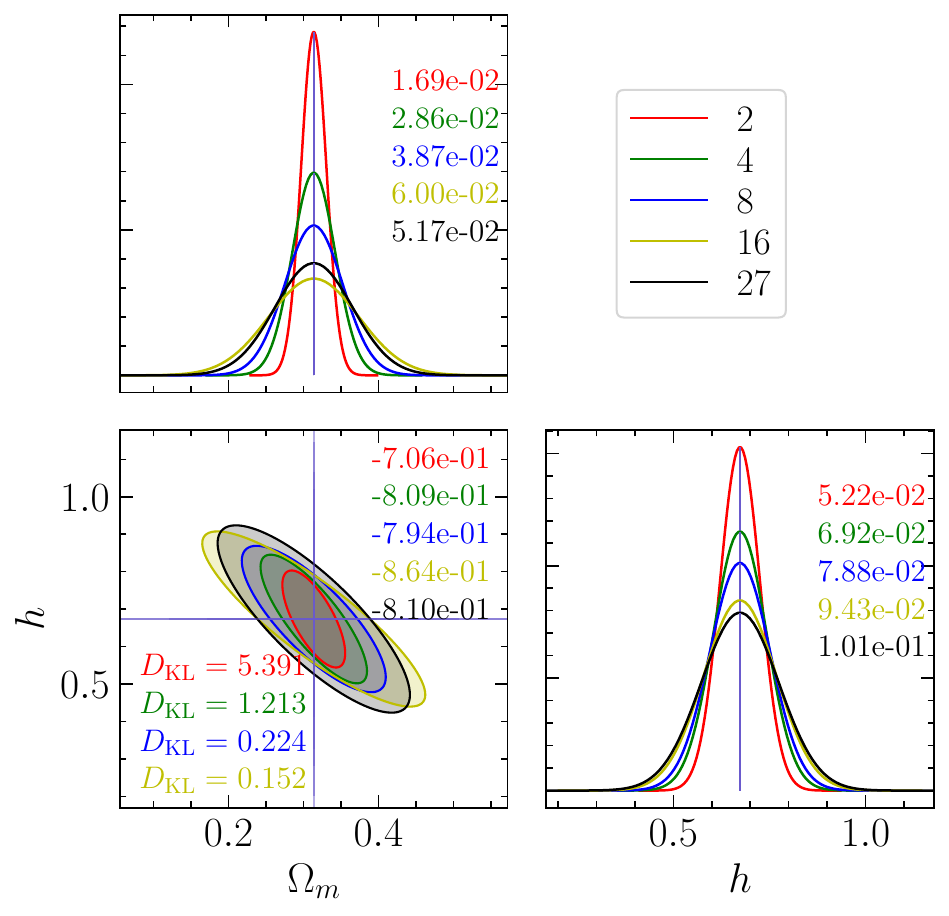}
         \caption{$n=2.0 \times 10^{-2}\, (\rm{Mpc})^{-3}$}
         \label{fig:corner_n2}
     \end{subfigure}
     \hfill
     \begin{subfigure}[b]{0.49\textwidth}
         \centering
         \includegraphics[width=\textwidth]{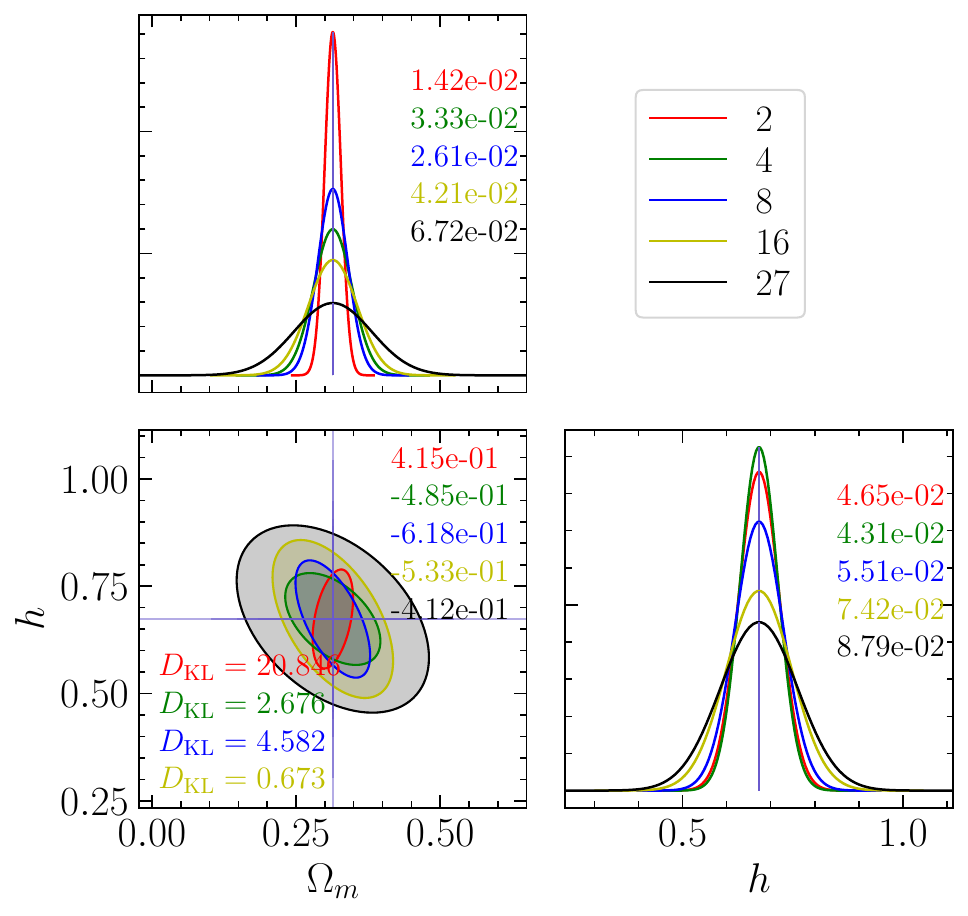}
         \caption{$n=2 \times 10^{-3}\, (\rm{Mpc})^{-3}$ }
         \label{fig:corner_n0.7}
     \end{subfigure}
\caption{Corner plots from Fisher analysis performed using the optimized percentiles for the two tracer samples of \Sahyadri\, using VVF, for \Om, $h$. }
        \label{fig:corner}
\end{figure}

\begin{figure}[h!]
     \centering
          \begin{subfigure}[b]{0.49\textwidth}
         \centering
         \includegraphics[width=\textwidth]{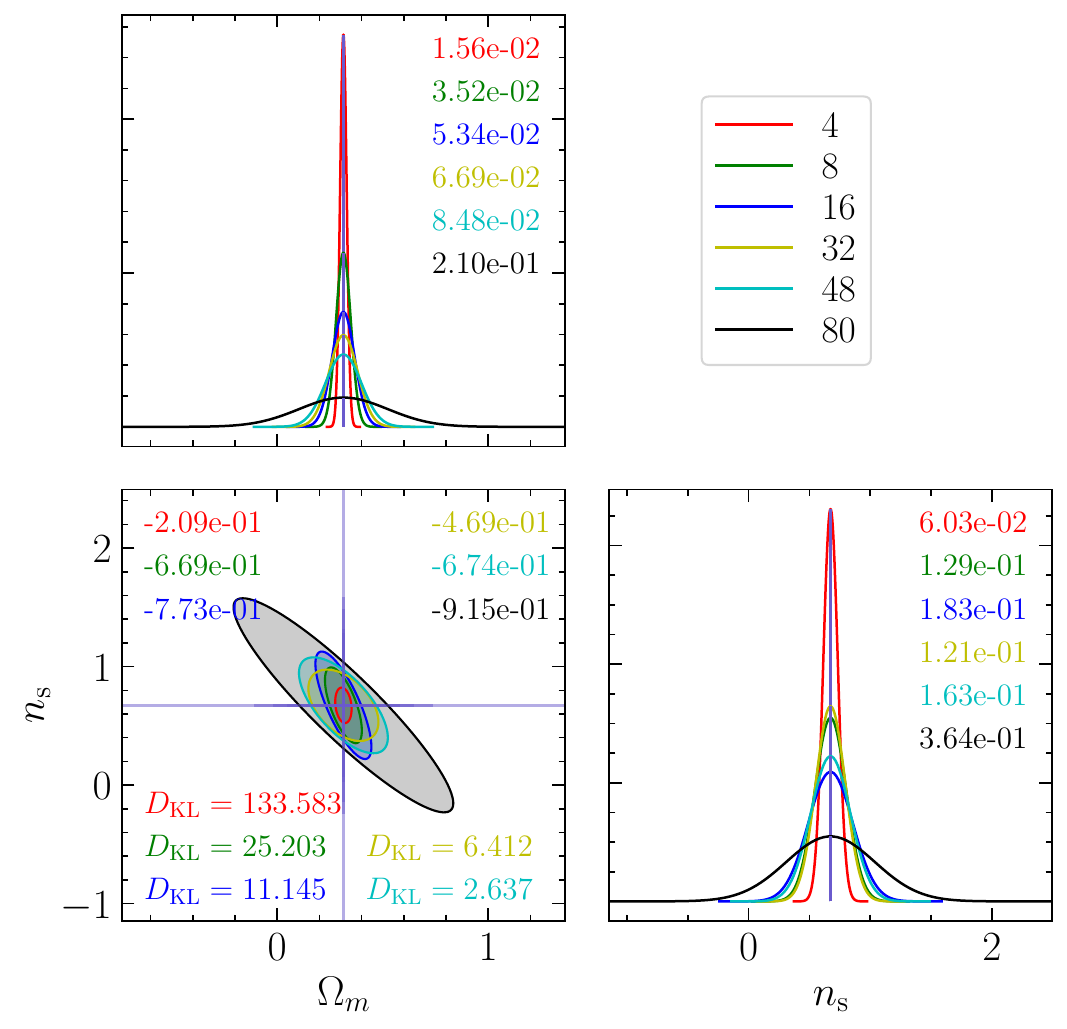}
         \caption{Full VVF, Unoptimized}
         \label{fig:corner_n0.7}
         \end{subfigure}
         \hfill
     \begin{subfigure}[b]{0.49\textwidth}
         \centering
         \includegraphics[width=\textwidth]{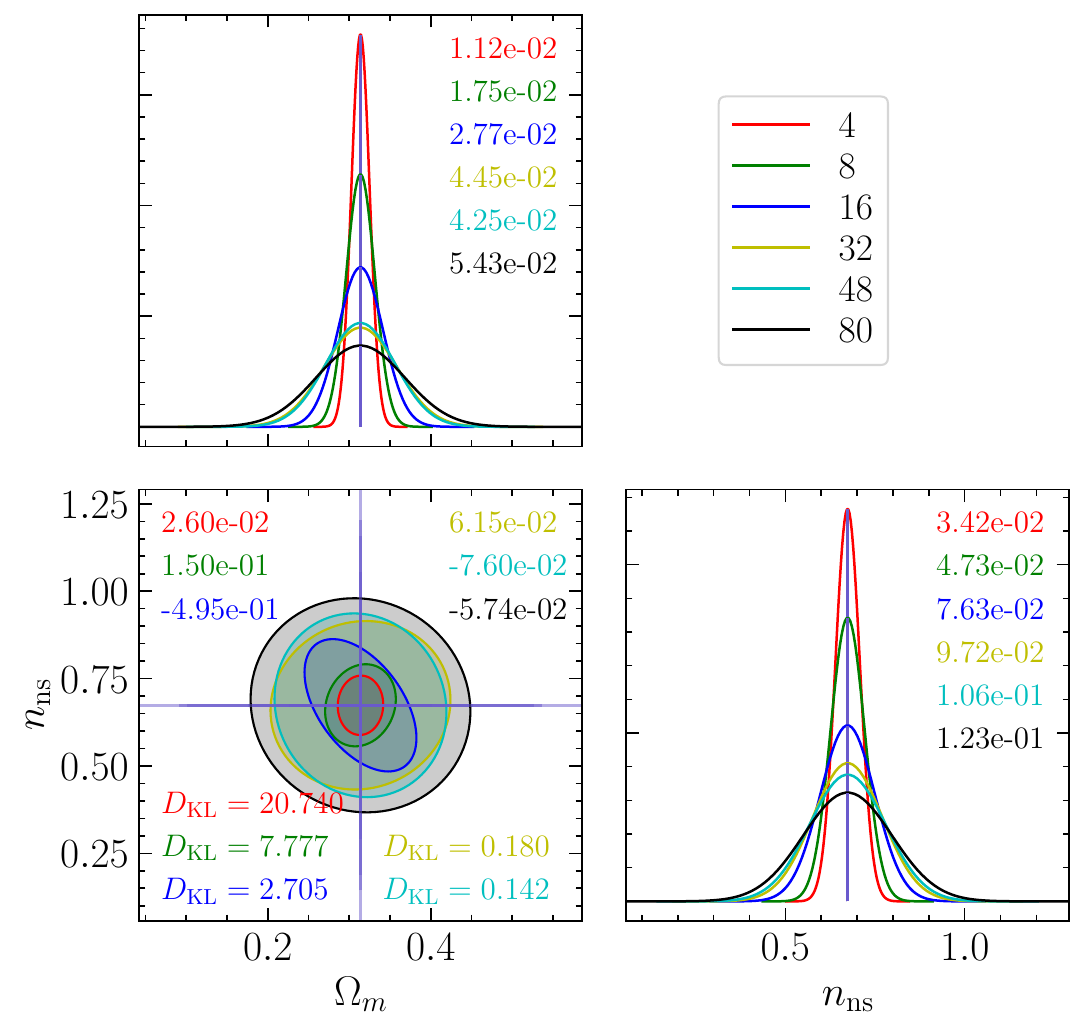}
         \caption{Random sub-sampled VVF, Optimized}
         \label{fig:corner_n2}
     \end{subfigure}
\caption{Corner plots from Fisher analysis performed using the \Sinhagad\, tracer sample ($n=2\times 10^{-4}\rm{Mpc}^{-3}$), for \Om, \ns, similar to Figure~\ref{fig:corner}. The unoptimized results by fusing the full VVF are given on the left, and the optimized version with random sub-sampling on the right. It is seen that optimization increases both the stability and constraining power.}
        \label{fig:VVF_sinhagad}
\end{figure}

\begin{figure}
    \centering
    \includegraphics[width=\linewidth]{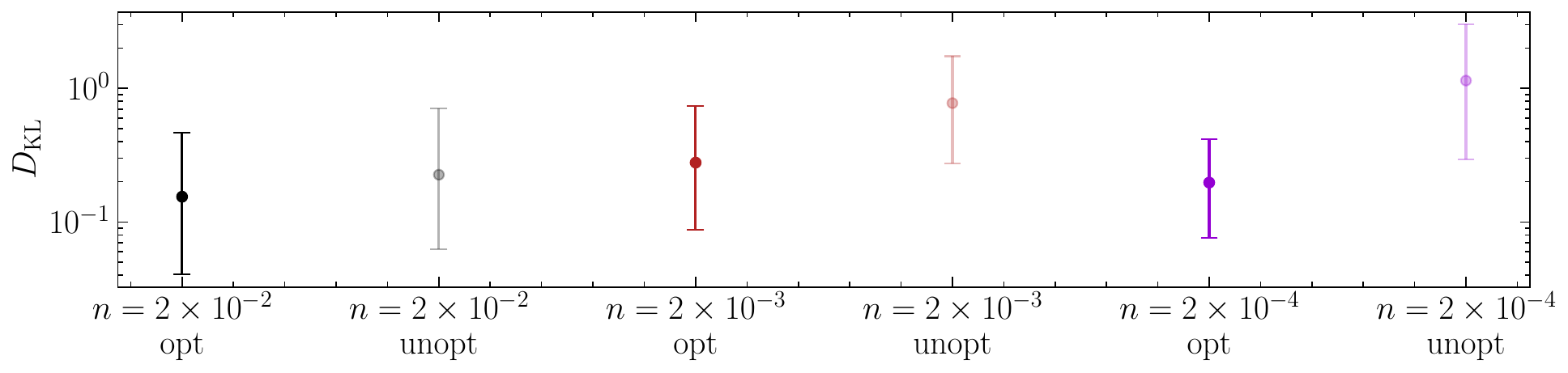}
    \caption{Comparison between Fisher forecasts using full VVF, unoptimized data vs random sub-sampled VVF, optimized data. The plot is similar to Figure~\ref{fig:mf_kl}, except for the lowest number density, where the averaging of derivatives has been done over $48$ realizations.}
    \label{fig:vvf_kl}
\end{figure}

Figure \ref{fig:corner} shows the Fisher constraints obtained using the optimized percentiles for both \Sahyadri\ samples, based on random sub-samples of the VVF. These forecasts are less well converged than the HMF results, though the highest–number-density sample shows clear signs of approaching convergence. For the lower–number-density sample, however, additional realizations are required to achieve stable and reliable constraints.

For the \Sinhagad\ suite, the availability of $800$ equivalent realizations for estimating the covariance matrix relaxes the restrictions on the number of usable VVF percentiles compared to \Sahyadri. After discarding percentiles separated by less than $2\%$, we retain $13$ percentiles for the Fisher analysis. The resulting constraints, shown in Figure \ref{fig:VVF_sinhagad}, demonstrate substantial improvement: although some discrepancy remains between the derivatives averaged over $48$ versus $80$ realizations, the overall convergence is much better, and the constraints shrink by factors of $\sim 3$–$4$ for $\ns$ and $\Om$ respectively. 

Finally, Figure \ref{fig:vvf_kl} presents the KL divergence between the accurate constraints and those obtained using 16 (\Sahyadri) or 48 (\Sinhagad) realizations at a time. In both cases, the optimization procedure leads to markedly lower $D_{\rm KL}$ values, clearly indicating enhanced robustness and stability of the Fisher forecasts.

 We find that the VVF can constrain  $\Om, \ns$ and $h$ at the level of $\leq 15\%$.  However, the absolute values of these constraints should not be over-interpreted: the tracer samples used here are not constructed to mimic realistic galaxy populations, and no variation in galaxy evolution physics has been included. Also, note that although \Sahyadri\ tracer samples have substantially higher number densities, the constraints on the cosmological parameters are comparable to those obtained from the \Sinhagad\ tracer sample. This is probably because the sub-volumes used in \Sinhagad\ are larger by a factor $27/8$, and some of the information is lost when using smaller sub-volumes.

\section{Discussion}
\label{sec:Discussion}
Fisher forecasts are a standard tool for quantifying the information content of cosmological summary statistics. However, their reliability can be compromised when derivative estimates are noisy because the number of realizations is limited, as is often the case in simulation-based studies. In such scenarios, using randomly chosen data points can lead to unstable constraints. 
In this work, we presented a formalism to identify informative and stable subsets of summaries of halo-based statistics, which maximize the stability and information content in the Fisher forecast using these statistics. We illustrated the use of this formalism with the Voronoi volume function (VVF, \cite{VVF2020}) and the halo mass function (HMF). Our method is motivated by practical limitations encountered in simulation-based cosmology, where both the number of realizations and the availability of cosmological variations are often constrained. Through this work, we offer a framework that remains effective across both well-sampled and data-limited regimes.

The formalism involves a two-step approach: averaging the statistic across multiple random sub-samples of the tracer population, followed by an optimization over subsets of summary statistic values (e.g., VVF percentiles). Random sub-sampling plays an important role in suppressing noise when the statistic is inherently noisy, as is the case when low number density tracers are used, while the subsequent optimization ensures that the selected data points contribute maximal information in a stable manner. The combination leads to tighter and more consistent parameter constraints compared to fixed, heuristic choices.

In the case of the low-resolution \Sinhagad\, simulation suite, we had access to multiple realizations per cosmology, allowing us to estimate the covariance matrix robustly and compute an accurate reference slope. We also split each box into sub-boxes, creating a large number of equivalent realizations. This enabled the definition of three metrics—$X$, $Y$, and $Z$—to capture the accuracy, informativeness, and stability of derivative estimates. Using these diagnostics, we select an optimized subset of points that yield consistent and strong constraints. This is illustrated using the VVF to constrain the cosmological parameters $\Om$ and $\ns$. Our analysis shows that these optimized subsets lead to comparatively stable Fisher forecasts compared to those obtained using un-optimized data points, with a median KL divergence of 0.076 bits with respect to the accurate constraints, when using $48$ equivalent realizations at a time.

In contrast, the high-resolution \Sahyadri\, suite represents a more challenging setting where only a single realization is available per cosmology. We also split the \Sahyadri\, boxes into 27 sub-boxes, which enabled us to estimate the covariance matrix as well as more robust derivative estimates. The optimization using halo mass function gave excellent results, with complete convergence in Fisher forecasts obtained with using less than $4-8$ sub-boxes at a time, depending on the  tracer sample. A very promising improvement was seen in both the size as well as stability of the constraints after applying the optimization procedure. The VVF-based constraints also showed promise, although the stability was not as good as for the mass function.
Although this work has focused on the VVF and HMF, the proposed method is general and can be applied to any summary statistic where derivative estimates are noisy. In practical analyses involving cosmological simulations or observational data, computational limitations and sparse sampling are common challenges. E.g., currently popular machine learning techniques such as SBI applications often require $\Cal{O}(100)$ realizations for robust derivative estimation, which might not always be feasible. Our formalism remains effective under such constraints and improves the reliability of parameter forecasts even with limited inputs.

Future work will focus on extending this framework in several directions. One priority is to develop a systematic approach to estimate the covariance in the high resolution simulations without needing to split it into multiple sub-boxes. If a covariance estimate is available, one can do with smaller number of sub-boxes for the derivative estimate, which will cut down on the cosmic variance, probably giving convergence with smaller number of necessary realizations. In addition,  we plan to test the method on combinations of multiple summary statistics -- such as the VVF and the two point correlation function -- to evaluate potential gains in parameter sensitivity. These improvements will allow for thorough testing of the method’s effectiveness and applicability in real-world cosmological applications.

\section*{Data availability}
The simulation products used in this work will eventually be made publicly available.

\acknowledgments
We gratefully acknowledge the participants of the Pune-Mumbai Cosmology \& Astro-Particle (PMCAP) series of meetings ( \url{https://www.tifr.res.in/~shadab.alam/PM_CAP_meeting/}) for stimulating discussions that inspired the construction of the simulations used in this work. We are especially grateful to Arka Banerjee for originally pointing out that halo-based statistics are prone to strong biases due to noisy derivatives. The research of AP is supported by the Associates Scheme of ICTP, Trieste.
This work made extensive use of the open source computing packages NumPy \citep{vanderwalt-numpy},\footnote{\url{http://www.numpy.org}} SciPy \citep{scipy},\footnote{\url{http://www.scipy.org}} Pandas \citep{mckinney-proc-scipy-2010, reback2020pandas},\footnote{\url{https://pandas.pydata.org}} Matplotlib \citep{hunter07_matplotlib},\footnote{\url{https://matplotlib.org/}} and Jupyter Notebook.\footnote{\url{https://jupyter.org}} The analysis was performed on the Pegasus cluster at IUCAA, Pune.\footnote{\url{http://hpc.iucaa.in}}

\bibliography{references}

\appendix

\section{Covariance and Fisher matrices}
\label{app:cov_fisher}
The covariance matrices used in the main text are estimated using sub-boxes in the default simulation realization(s) treated as $N$ independent realizations. The covariance is computed using the standard formula:
\begin{equation}
    C_{ij}= \frac{1}{N-1}\sum_{k=2}^{N+1} (y_i^k- \langle y_i \rangle)(y_j^k- \langle y_j \rangle).
\end{equation}
Here, $y_i^k$ is the value of the statistic at $i$-th 
bin or percentile, for the $k$-th realization, $\langle y_i \rangle \equiv (1/N)\sum_{k=2}^{N+1}y_i^k$. \\
When only $\pm \Delta$ variations of the cosmological parameter are available, as in case of the high resolution \Sahyadri\, simulations, the derivatives with respect to the cosmological parameters are estimated using the three-point 
central difference method across seed-matched simulations:
\begin{equation}
    \tilde{y}_{,\alpha} \equiv \frac{\partial \tilde{y}}{\partial \theta_\alpha}= \frac{\tilde{y}(\theta_{0\alpha}+ \Delta\theta_\alpha)-\tilde{y}(\theta_{0\alpha}- \Delta\theta_\alpha)}{2\Delta \theta_\alpha}
\end{equation}
where the data vectors $\tilde{y}$ are calculated from the $\pm \Delta$ variations of the $\alpha$th cosmological parameter, whose fiducial value is $\theta_{0\alpha}$.\\
When $\pm 2\Delta, \pm \Delta, \pm 0.5\Delta$ variations are available as is the case for \Sinhagad, we fit a 3rd order polynomial through the 7 points and use the 1st coefficient as an estimate of the slope.\\
While performing the Fisher analysis, we make the following assumptions: (a) the likelihood is Gaussian, and (b) the data covariance matrix $C_{ij}$ is independent of the cosmological parameters $\theta_\alpha$ within the $\Delta \theta$ variations considered here. Then, the Fisher matrix simplifies to \citep{Tegmark+1997}:
\begin{equation}
    F_{\alpha\beta} =\frac{\partial \tilde{y}}{\partial \theta_\alpha}C^{-1}\frac{\partial\tilde{y}}{\partial\theta_\beta}.
\end{equation}
The inverse Fisher matrix $F^{-1}_{\alpha \beta}$ corresponds to the posterior covariance matrix of the cosmological parameters under the assumption of very broad priors.

\section{Effect of random sub-sampling}
\label{App: bootstrp vs full VVF}
\begin{table}[h!]
    \centering
    \begin{tabular}{|c|c|c|}
    \hline
        Number density ($\rm{Mpc}^{-3})$ &    \multicolumn{2}{c|}{Number of valid points}  \\
        \cline{2-3} 
        & With random sub-sampling & Without random sub-sampling\\
        
        \hline
        $2 \times 10^{-2}$ & 55 & 44 \\
        \hline
        $2 \times 10^{-3}$& 28 & 31\\
        \hline
        $2 \times 10^{-4}$ &26 &0\\
        \hline
    \end{tabular}
    \caption{Number of valid points available for different parameter variations for the \Sinhagad\, tracer sample.}
    \label{tab:tab_bootstrap}
\end{table}

Here, we compare the quality of the VVF estimated using the full sample versus the random sub-sampled version, with the aim of determining which approach is more suitable for performing Fisher forecasts. To make this comparison, we apply the analysis described in Section~\ref{sec:Statistics}, selecting valid data points based on the metrics $X, Y, Z$ defined earlier. The same cuts in $X, Y, Z$ that reported in the paper are used for the full sample as well.

Table~\ref{tab:tab_bootstrap} summarizes the number of valid points retained for each tracer sample under both methods. It is seen that the random sub-sampled VVF results in a greater number of valid data points, especially for the lowest number density sample. For the middle number density, the full VVF gives slightly, but not substantially larger number of valid points. This indicates that random sub-sampling improves the stability of the derivative estimates and provides a broader pool of usable points for Fisher analysis.  We therefore adopt the random sub-sampled VVF as the default estimator in our study.

\begin{figure}[h!]

     \centering
     \begin{subfigure}[b]{0.49\textwidth}
         \centering
         \includegraphics[width=\textwidth]{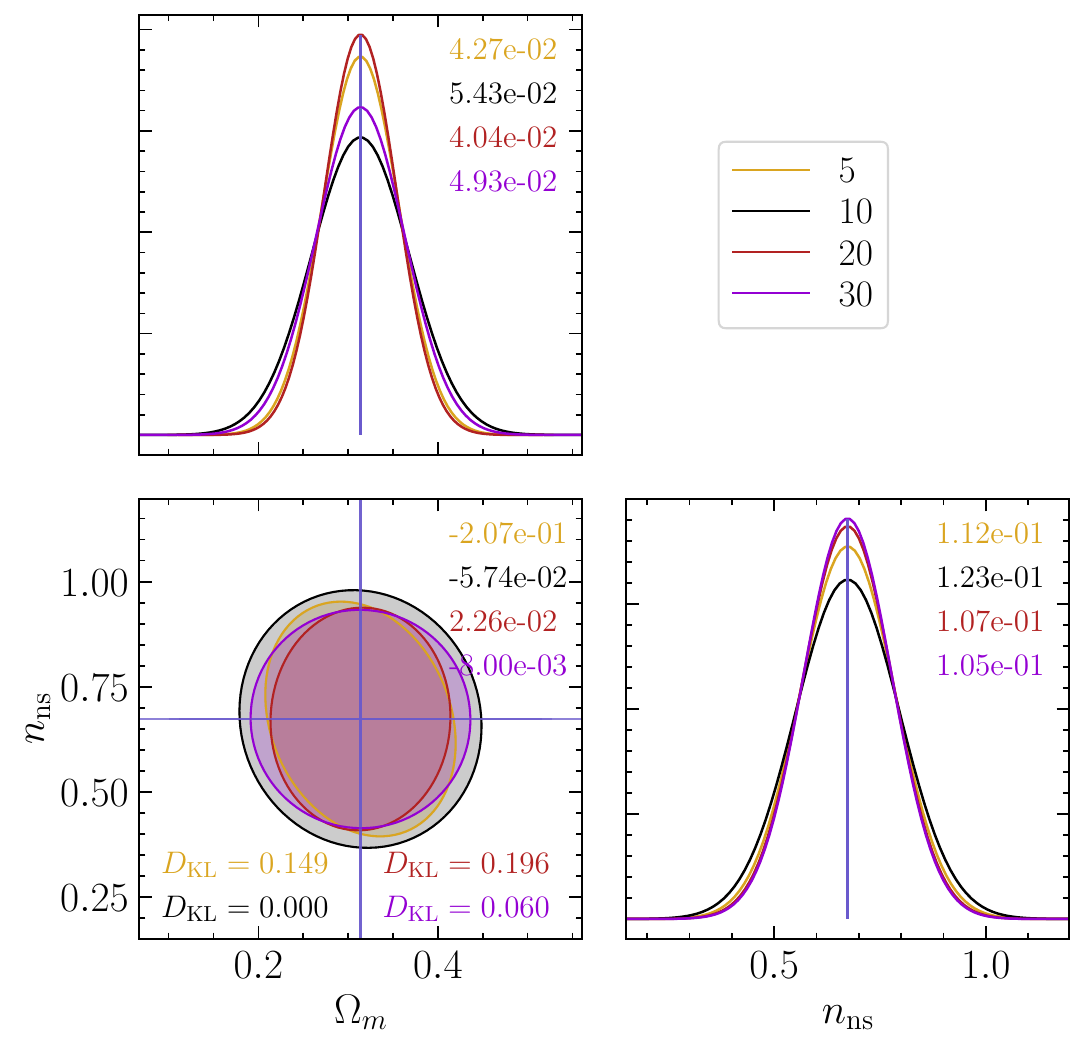}
     \end{subfigure}
     \hfill
     \begin{subfigure}[b]{0.49\textwidth}
         \centering
         \includegraphics[width=\textwidth]{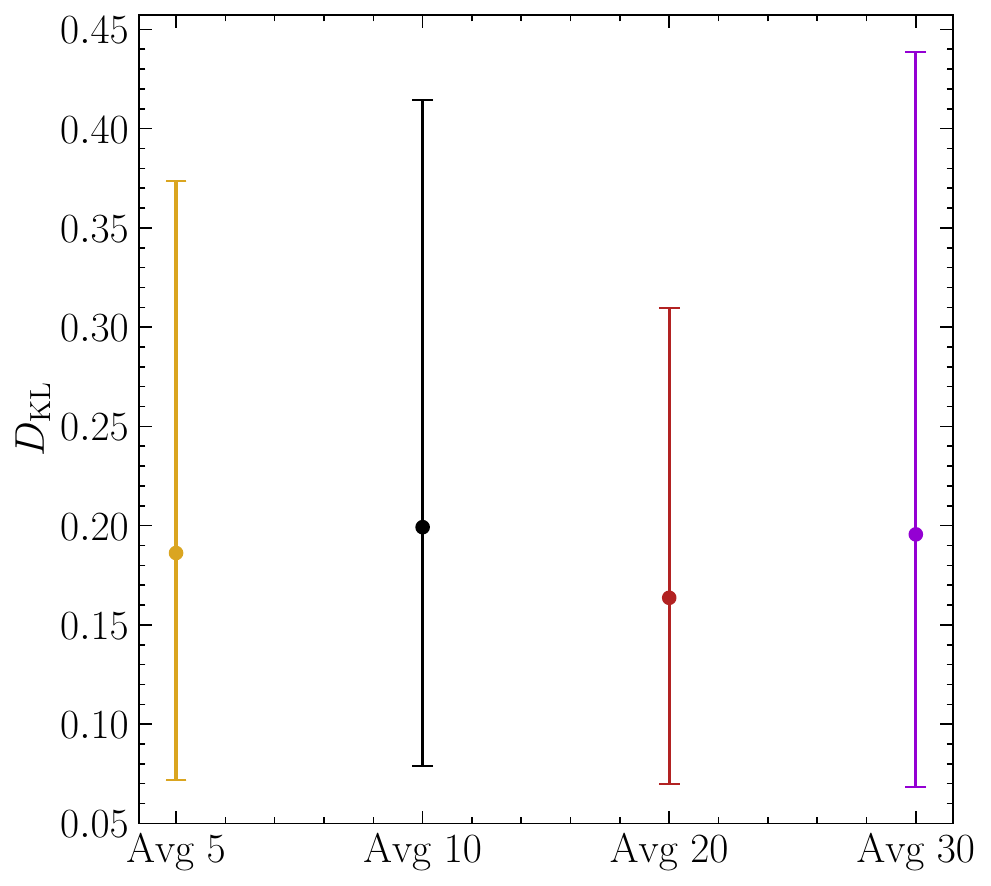}
     \end{subfigure}
    \caption{Comparison of forecasts obtained using different random sub-sampling averages for the \Sinhagad\, tracer sample. 
    \emph{Left panel:} Final Fisher forecasts using derivatives averaged over all $80$ realizations, with percentile selections optimized separately for each averaging choice. The KL divergence is evaluated with respect to the default choice of averaging over $10$ sub-samples. 
    \emph{Right panel:} Stability of the forecasts, quantified by the KL divergence between results using derivatives averaged over $48$ and $80$ realizations. The forecasts are seen to be relatively insensitive to the choice of sub-sampling average.}
    \label{fig:bs_sensitivity}
    \end{figure}

To test the robustness of our random sub-sampling procedure, we construct the VVF by averaging over $5$, $20$, and $30$ random sub-sampled realizations (each using $70\%$ of the \Sinhagad, tracer set), in addition to the default choice of averaging over $10$ realizations reported in the main text. We repeat the full optimization pipeline and obtain Fisher forecasts using derivatives averaged over all $80$ realizations. For averages of $10$, $20$, and $30$, the original $X, Y, Z$ thresholds remain valid, whereas for the case with only $5$ averages we relax $Z_{\rm th}$ to $1.5$. The results, shown in Figure~\ref{fig:bs_sensitivity}, demonstrate that the optimized Fisher forecasts are largely insensitive to the number of random sub-sampling averages.

\section{Choice of parameter variations}
\label{App: parameter variations}
\begin{table}[h!]
    \centering
    \begin{tabular}{|c|c|}
    \hline
        Variation & Number of valid points  \\
        \hline
        $\Delta$ & 26 \\
        \hline
         $\Delta/2$ & 14 \\
        \hline
    \end{tabular}
    \caption{Number of valid points available for different parameter variations for the \Sinhagad\, tracer sample.}
    \label{tab:tab_appendix}
\end{table}
In many practical settings, the computational cost of running high-resolution simulations limits the number of cosmological parameter variations that can be explored. This constraint applies to our under-construction \Sahyadri\ simulation suite, where we restrict ourselves to $\pm \Delta$ variations around the fiducial values for each cosmological parameter. On the other hand, the lower-resolution simulations in the \Sinhagad\, suite are computationally inexpensive, making them ideal for testing different variation schemes and assessing their impact on the stability and constraining power of derivative estimates. We use this flexibility to compare the performance of two choices: $\pm \Delta/2$ and $\pm \Delta$, where $\Delta$ corresponds to the default parameter step size adopted in this paper.

In this paper, derivatives are computed using a three-point central difference method. Choosing $\Delta$ to be too large risks violating the assumption of local linearity in the statistic–parameter relation. Conversely, too small a variation increases the noise arising from stochastic effects such as halo finder fluctuations, resulting in highly unstable derivative estimates.

Following the same procedure as in Appendix~\ref{App: bootstrp vs full VVF}, we evaluate the metrics $X, Y, Z$ for both variation schemes. The $\Delta$ variation leads to lower (and more favorable) values of these diagnostics, resulting in a larger number of points being accepted as valid, as shown the summary in Table~\ref{tab:tab_appendix}. We then perform the Fisher analysis using the optimized percentile selections for both schemes and compute the KL divergence between the Fisher matrices obtained using derivatives averaged over 48 realizations versus those averaged over 80 realizations. For $\pm \Delta/2$ variations this is $0.435_{-0.334}^{+0.739}$, substantially larger than the value for the $\pm \Delta$ variation: $0.203_{-0.125}^{+0.214}$. Based on these results, we adopt the $\pm \Delta$ variation scheme for the remainder of this paper.

\section{Sensitivity to the order of the polynomial fit}
\label{App: Polyfit}

\begin{figure}[h!]
     \centering
     \begin{subfigure}[b]{0.49\textwidth}
         \centering
         \includegraphics[width=\textwidth]{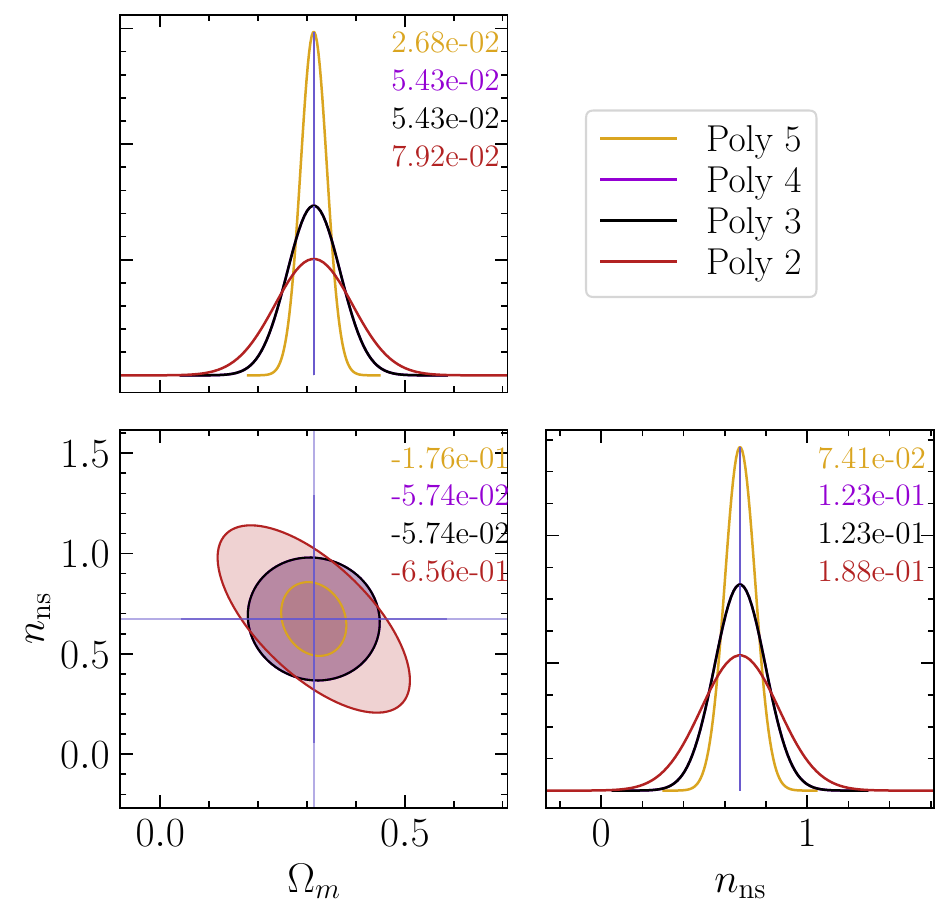}
         \label{fig:corner_poly}
     \end{subfigure}
     \hfill
     \begin{subfigure}[b]{0.49\textwidth}
         \centering
         \includegraphics[width=\textwidth]{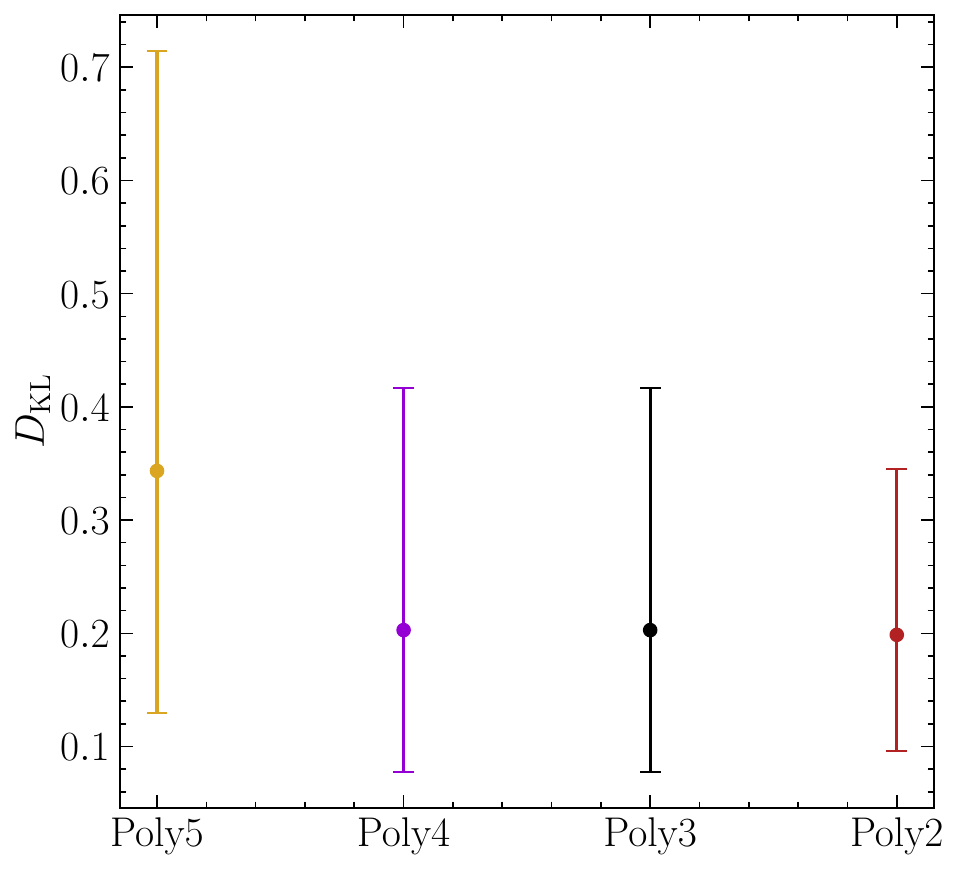}
         \label{fig:corner_kl_poly}
     \end{subfigure}
\caption{Comparison of forecasts obtained using different polynomial orders for derivative estimation for the \Sinhagad\, tracer sample.
\emph{Left panel:} Final Fisher forecasts, using derivatives averaged over all $80$ realizations and percentile selections optimized separately for each fit order.
\emph{Right panel:} Stability of the forecasts, quantified by the KL divergence between results using derivatives averaged over $48$ and $80$ realizations.
Among the tested options, the third–order polynomial provides the most stable and reliable performance.}
    \label{fig:poly_sensitivity}
\end{figure}

To obtain accurate derivative estimates in the \Sinhagad–like setting, we fit a third–order polynomial to the seven available points and take its linear coefficient as the derivative. To check the sensitivity of our results to this choice, we repeat the analysis using polynomial fits of order $2, 4$ and $5$. Figure~\ref{fig:poly_sensitivity} compares the resulting Fisher forecasts, and also shows the stability quantified through the KL divergence between forecasts obtained by averaging derivatives over 48 vs 80 realizations (the latter treated as the truth).

Fitting a fourth–order polynomial yields identical optimized percentile selections and Fisher forecasts relative to the third–order case, demonstrating robustness to moderate changes in fit order. A second–order fit, on the other hand, identifies a larger number of valid points and leads to weaker Fisher constraints, although the stability remains comparable to the third–order result. A fifth–order fit represents overfitting. For fits of order $2-4$ the original $X, Y, Z$ constraints are retained. In contrast, the fifth–order fit yields zero valid points for this choice. Therefore, we relax $Z_{\rm{th}}$ to a higher value, $3$. Although this adjustment produces tighter Fisher constraints, the gain comes with a clear loss in stability as seen from the left panel in the figure. Taken together, these tests indicate that while modest changes in the polynomial order do not affect our conclusions, the third-order fit offers the most reliable performance and is therefore used throughout.
\end{document}